\documentclass{appolb}
\usepackage{epsfig}

\begin{document}
\title{Relativistic Noise
\thanks{Presented at Strangeness in Quark Matter 2011, Krakow, Poland.}%
}
\author{Joseph Kapusta
\address{School of Physics \& Astronomy, University of Minnesota, Minneapolis, MN 55455,USA}
\\
\vskip6pt
 in collaboration with\\
\vskip6pt
Berndt M\"uller
\address{Department of Physics, Duke University, Durham, NC 27708-0305, USA}
\and
Misha Stephanov
\address{Department of Physics, University of Illinois, Chicago, IL 60607, USA}
}
\maketitle
\begin{abstract}
The relativistic theory of hydrodynamic fluctuations, or noise, is derived and applied to high energy heavy ion collisions.  These fluctuations are inherent in any space-time varying system and are in addition to initial state fluctuations.  We illustrate the effects with the boost-invariant Bjorken solution to the hydrodynamic equations. Long range correlations in rapidity are induced by propagation of sound modes.  The magnitude of these correlations are directly proportional to the viscosities.  These fluctuations should be enhanced near a phase transition or rapid crossover. 
\end{abstract}
  
\section{Introduction}

Cosmology has experienced tremendous advances in the past decade.  These advances have been driven by observations.  Much of the information comes from fluctuations in temperature of the cosmic microwave background radiation (CMBR) as observed by the Wilkinson Microwave Anisotropy Probe (WMAP) \cite{WMAP}.   The results are usually expressed in terms of the power spectrum, $l(l+1) C_l/2\pi$ versus the multipole moment $l$, where
\begin{eqnarray}
T(\theta,\phi) &=& \sum_{lm} a_{lm} Y_{lm}(\theta,\phi) \nonumber \\
C_l &=& \langle |a_{lm}|^2 \rangle
\end{eqnarray}
and where the averaging is done over points in the sky.  In WMAP7 the power spectrum extends up to values of $l$ on the order of 1000.  At the Relativistic Heavy Ion Collider (RHIC) and at the Large Hadron Collider (LHC) jets clearly stand out against a background of low transverse momentum particles in a lego plot.  In Pb-Pb collisions at the LHC these jets can have transverse momentum as high as 100 to 150 GeV/c compared to the background particles of several GeV/c \cite{jets}.  Here we are not interested in the jets but in the fluctuations in the background.  What information can we extract from these fluctuations?  

The theory behind fluctuations in the CMBR is highly nontrivial \cite{Dodelson} even if the basic ideas are rather intuitive.  One should expect a similar nontriviality in the theory for high energy heavy ion collisions.  There are at least four important sources of fluctuations in such collisions:
\begin{itemize}
\setlength{\itemsep}{0pt}
\item[(i)] {\em Initial state fluctuations:}  These arise because nuclei are composed of nucleons which in turn are composed of quarks and gluons.  The fluctuations are both statistical and quantum in nature.

\item[(ii)] {\em Hydrodynamic fluctuations:}  These arise due to finite particle number fluctuations in a given coarse-grained fluid cell.

\item[(iii)] {\em Fluctuations induced by jets:}  Jet production is a random process.  They deposit energy and momentum as they traverse the system.

\item[(iv)] {\em Freeze-out fluctuations:}  These arise when converting a coarse-grained fluid cell into individual particles which are subsequently described by a Monte Carlo transport model. 
\end{itemize} 
The goal here is to develop a relativistic theory of hydrodynamic fluctuations for application to high energy heavy collisions.  After describing the basic theory we will apply it to the boost-invariant solution of Bjorken for high energy collisions.  Although not realistic enough to compare directly with data it will demonstrate that correlations develop whose magnitude and shape are controlled, in a quantitative manner, by the shear and bulk viscosities as well as by the equation of state.

Intuitively, hydrodynamic fluctuations or noise becomes important when gradients of temperature, density, and particle composition become large.  In such situations the coarse-grained fluid cells must be relatively small to adequately represent these gradients.  The smaller the cell the more important fluctuations will be.  Certainly this is the case with high energy heavy ion collisions.  Analogous situations may be found throughout the physical, chemical and biological literature; see \cite{Donev} and citations within.  For example, Eggers \cite{Eggers} performed a theoretical study of the breakup of liquid nanojets with the conclusion that ``noise is the driving force behind pinching, speeding up the breakup to make surface tension irrelevant".  Similar conclusions were reached by Kang and Landman \cite{Kang} who studied the breakup of liquid nanobridges with a molecular dynamics approach, with a lubrication equation (smooth fluid dynamics), and with a stochastic lubrication equation.  Inclusion of noise in the lubrication equation provided results very similar to the molecular dynamics simulations.

With these as motivations we proceed to the general study of relativistic hydrodynamic fluctuations in the next section followed by an application to the boost invariant hydrydynamics of Bjorken.

\section{Relativistic Hydrodynamic Fluctuations}

Now we turn to the topic of hydrodynamic fluctuations.  The energy-momentum tensor density for a perfect fluid is 
\begin{equation}
T^{\mu\nu}_{\rm ideal} = - Pg^{\mu\nu}+wu^{\mu}u^{\nu} \, .
\end{equation}
Here $w=P+\epsilon=Ts + \mu n$ is the local enthalpy density, $\mu$ is the baryon chemical potential, $n$ is the baryon density, and $u^{\mu}$ is the local flow velocity.  The metric is 
$(+,-,-,-)$.  Corrections to this expression are proportional to first derivatives of the local quantities whose coefficients are the shear viscosity $\eta$, bulk viscosity $\zeta$, and thermal conductivity $\chi$.  Explicit expressions may be found in textbooks \cite{fluid1,fluid2} which are useful to summarize here.  Dissipative contributions are added to the energy-momentum tensor and baryon current as follows.
\begin{eqnarray}
T^{\mu\nu} &=& T^{\mu\nu}_{\rm ideal} + \Delta T^{\mu\nu}
\nonumber \\
J^{\mu} &=& n u^{\mu} + \Delta J^{\mu}
\end{eqnarray}
There are two common definitions of the flow velocity in relativistic dissipative fluid dynamics which are important to distinguish.  In the Landau-Lifshitz approach $u^{\mu}$ is the velocity of energy transport. In the Eckart approach $u^{\mu}$ is the velocity of baryon number flow.  In high energy heavy ion collisions, at the upper range of RHIC energies and at the LHC, the net baryon number is very small compared to the entropy density or to the number of baryons plus anti-baryons.  Therefore the Landau-Lifshitz approach is the relevant one.

In the Landau-Lifshitz approach the dissipative part of the energy-momentum tensor satisfies $u_{\mu} \Delta T^{\mu\nu} = 0$.  The most general form of the energy-momentum tensor is
\begin{equation}
\Delta T^{\mu\nu} = 
\eta \left(\Delta^{\mu} u^{\nu} + \Delta^{\nu} u^{\mu}\right)
+\left({\textstyle{\frac{2}{3}}} \eta - \zeta\right) H^{\mu\nu} \partial \cdot u \, .
\label{LLvis}
\end{equation}
Here
\begin{equation}
H^{\mu\nu} = u^{\mu} u^{\nu} - g^{\mu\nu}
\end{equation}
is a projection tensor normal to $u^{\mu}$,
\begin{equation}
\Delta_{\mu} = \partial_{\mu} - u_{\mu} \left( u \cdot \partial \right)
\end{equation}
is a derivative normal to $u^{\mu}$, and
\begin{equation}
Q_{\alpha} = \partial_{\alpha} T - T \left( u \cdot \partial \right) u_{\alpha}
\end{equation}
is a heat flow vector whose nonrelativistic limit is ${\bf Q} = - \mbox{\boldmath $\nabla$} T$.   The baryon current is modified by
\begin{equation}
\Delta J^{\mu} = \chi \left(\frac{nT}{w}\right)^2 \Delta^{\mu} \left(\beta \mu \right) \, ,
\end{equation}
which satisfies $u_{\mu} \Delta J^{\mu} = 0$.  This insures that 
$n$ is the baryon density in the local rest frame.  The entropy current in this approach is
\begin{equation}
s^{\mu} = s u^{\mu} - \beta \mu \Delta J^{\mu} \, .
\end{equation}
In the local rest frame entropy is generated according to the divergence
\begin{eqnarray}
\partial_{\mu}s^{\mu} &=& \frac{\eta}{2T}
\left( \partial_iu^j + \partial_ju^i - {\textstyle{\frac{2}{3}}} \delta^{ij} \nabla \cdot {\bf u}
\right)^2 \nonumber \\
& & \mbox{} + \frac{\zeta}{T} \left( \nabla \cdot {\bf u} \right)^2
+ \frac{\chi}{T^2} \left( \nabla T + T \dot{{\bf u}}\right)^2 \, .
\end{eqnarray}
The term $T \dot{{\bf u}}$ is a relativistic correction to $\nabla T$, being smaller by a factor of $1/c^2$ in physical units.  All three dissipation coefficients must be non-negative to insure that entropy can never decrease.

Next we add small fluctuations to the energy-momentum tensor
\begin{equation}
T^{\mu\nu} = T^{\mu\nu}_{\rm ideal} + \Delta T^{\mu\nu}
 + S^{\mu\nu}
\end{equation}
and to the baryon current
\begin{equation}
J^{\mu} = n u^{\mu} + \Delta J^{\mu} + I^{\mu} \, .
\end{equation}
The fluctuations must satisfy the conditions $u_{\mu} S^{\mu\nu} = 0$ and $u_{\mu} I^{\mu} = 0$, which are the same conditions satisfied by $\Delta T^{\mu\nu}$ and by $\Delta J$.  These fluctuations have zero average value at every space-time point.  The averaging is done with an ensemble of heavy ion collisions, all prepared with exactly the same initial conditions when the hydrodynamic description may be applied.  However, the average of a product of fluctuations is not necessarily zero.  We follow section 88 on hydrodynamic fluctuations of \cite{statphys2} to derive these correlators.  After some analysis we find
\begin{displaymath}
\langle S^{\mu\nu}(x_1) S^{\alpha\beta}(x_2) \rangle =
2T \left[ \eta \left( H^{\mu\alpha} H^{\nu\beta} + H^{\mu\beta} H^{\nu\alpha}
\right) \right.
\end{displaymath}
\begin{equation}
+ \left. \left(\zeta - {\textstyle{\frac{2}{3}}} \eta \right) H^{\mu\nu} H^{\alpha\beta} \right]
\delta (x_1 - x_2)
\end{equation}
and
\begin{equation}
\langle I^{\mu}(x_1) I^{\nu}(x_2) \rangle = 2 \chi 
\left(\frac{nT}{w}\right)^2 H^{\mu\nu} \delta (x_1 - x_2) \, .
\end{equation}
The mixture $\langle S^{\mu\nu}(x_1) I^{\alpha}(x_2) \rangle$ naturally enough is zero.   These correlation functions have their origin in the fluctuation-dissipation theorem.  There are two essential observations concerning them.  First, they are proportional to Dirac delta-functions.  Different coarse-grained fluid cells are assumed to be independent.  Second, the magnitudes of the correlations are directly proportional to the shear and bulk viscosities and to the thermal conductivity.  This is where the microscopic physics lies.

The procedure for implementing these stochastic sources are as follows.\\
$\bullet$ Solve the hydrodynamic equations for an arbitrary source function.  One can imagine doing this in principle although it may be difficult in practice.\\
$\bullet$ Perform averaging using the above expressions for the sources to obtain observable correlation functions.\\
$\bullet$ The stochastic fluctuations may or may not be perturbative, depending on the physical conditions.\\
To develop an understanding of the implementation we consider a simple model for heavy ion collisions in the next section.

\section{Boost Invariant Model}

It is insightful to work out a particular example; we shall do this for the well-known boost-invariant Bjorken model.  In the absence of fluctuations the important results are that the temperature depends only on the proper time as $T=T(\tau)$, the entropy density decreases with proper time as $s(\tau)=s_0\tau_0/\tau$, where $\tau_0$ is the equilibration time, and the fluid velocity is $u^{\mu}=(\cosh\xi,0,0,\sinh\xi)$, where $\xi$ is the space-time rapidity.  Space-time fluctuations may be expressed as
\begin{eqnarray}
T &=& T(\tau) + \delta T(\xi,\tau) \nonumber \\
u^{\mu} &=& (\cosh(\xi+\omega(\xi,\tau)),0,0,\sinh(\xi+\omega(\xi,\tau))) \, .
\end{eqnarray}
The function $\omega$ is dimensionless, and it is convenient to also use the dimensionless variable $\rho \equiv \delta s/s$ instead of $\delta T$ (the latter are related by thermodynamic identities).  For purposes of illustration, these fluctuations are treated as perturbations.  After lengthy calculation one finds the typical linear response relations
\begin{equation}
\tilde{X}(k,\tau) = - \int_{\tau_0}^{\tau} \frac{d\tau'}{\tau'}
\tilde{G}_{X}(k;\tau,\tau') \tilde{f}(k,\tau')
\end{equation} 
where $X$ is either $\rho$ or $\omega$.  This relation is given in terms of the variable $k$ which corresponds to the Fourier transform of the variable $\xi$.  The function $f$ is the single scalar function representing the noise which can expressed as
\begin{equation}
S^{\mu\nu}=w(\tau) f(\xi,\tau) H^{\mu\nu} \, .
\label{hS}
\end{equation}
The correlation functions are
\begin{displaymath}
\langle X(\xi,\tau_f) Y(0,\tau_f) \rangle = \frac{1}{\pi A} \int_{\tau_0}^{\tau_f} \frac{d\tau}{\tau^3} \frac{T(\tau)}{w^2(\tau)} \left[ {\textstyle{\frac{4}{3}}} \eta(\tau)
+ \zeta(\tau) \right]
\end{displaymath}
\begin{equation}
\times \int_{-\infty}^{\infty} dk {\rm e}^{ik \xi} {\tilde G}_{XY}(k;\tau_f,\tau) \, .
\end{equation}
Here ${\tilde G}_{XY}(k;\tau_f,\tau)={\tilde G}_X(k;\tau_f,\tau) {\tilde G}_Y(-k;\tau_f,\tau)$, $A$ is the effective transverse area of the colliding nuclei, and $\tau_f$ is the freeze-out time at which the transition from hydrodynamic flow to free-streaming of particles takes place.  The Green functions ${\tilde G}_X(k,\tau,\tau')$ are even functions of $k$ and calculable.  

Take, for example, the equation of state $P = \frac{1}{3}\epsilon$, and treat the underlying expansion as essentially inviscid.  Then
\begin{equation}
{\tilde G}_{\rho}(k;\tau,\tau') = \left(\frac{\tau'}{\tau}\right)^{1/3} \left[ \frac{2+3\gamma - 9\gamma^2}{6\gamma} \left(\frac{\tau}{\tau'}\right)^{\gamma} - \frac{2-3\gamma - 9\gamma^2}{6\gamma} \left(\frac{\tau}{\tau'}\right)^{-\gamma}\right]
\end{equation}
where $\gamma=\frac{1}{3}\sqrt{1-3k^2}$.  Note that $\gamma$ may be real or imaginary.  Similar expressions can be written down for the other response functions.  The Fourier transformed functions are singular, with Dirac delta-functions and derivatives of them at $\xi=0$ and at the sound horizon $\xi = 2 v_s \ln(\tau/\tau')$.  Figure 1 shows the regular part of ${\tilde G}_{\rho\rho}$ while Fig. 2 shows the singular part which is smeared by a Gaussian of arbitrary with to display the singularities.  The origin of the singularities are the space-time delta-functions in the original correlation functions; this represents white noise.  It is possible to cure these singularities by using finite range correlations, which is colored noise.

\begin{figure}[t]
  \centering
\includegraphics[width=0.85\linewidth]{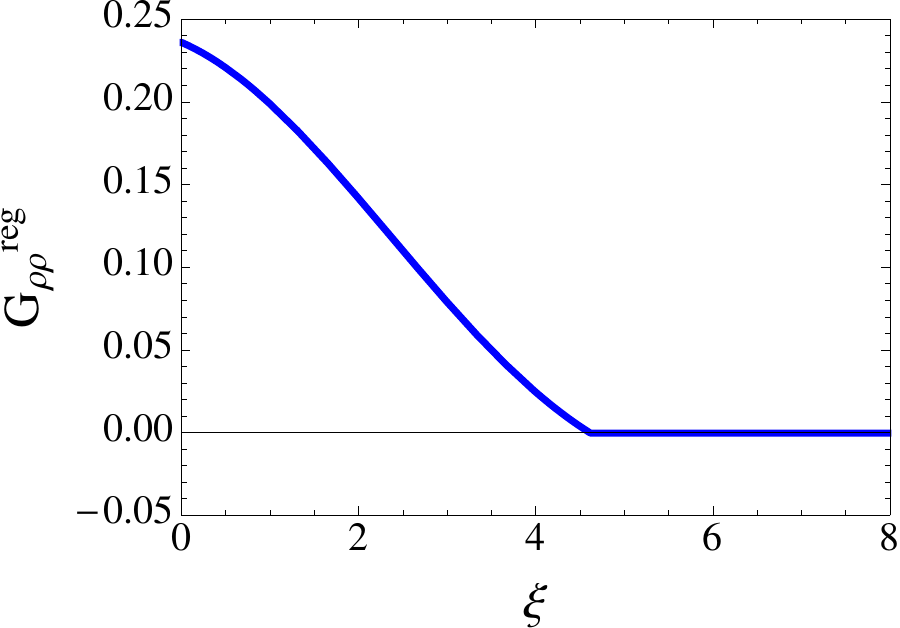}
  \caption{The regular (continuous) part of the correlator $G_{\rho\rho}(\xi;\tau_f,\tau)$ with $v_s^2=1/3$ and $\ln(\tau_f/\tau)=4$.  Note the sound horizon at $\xi=2v_s \ln(\tau_f/\tau)$.}
  \label{fig:G_reg}
\end{figure}
\begin{figure}[b]
  \centering
\includegraphics[width=0.85\linewidth]{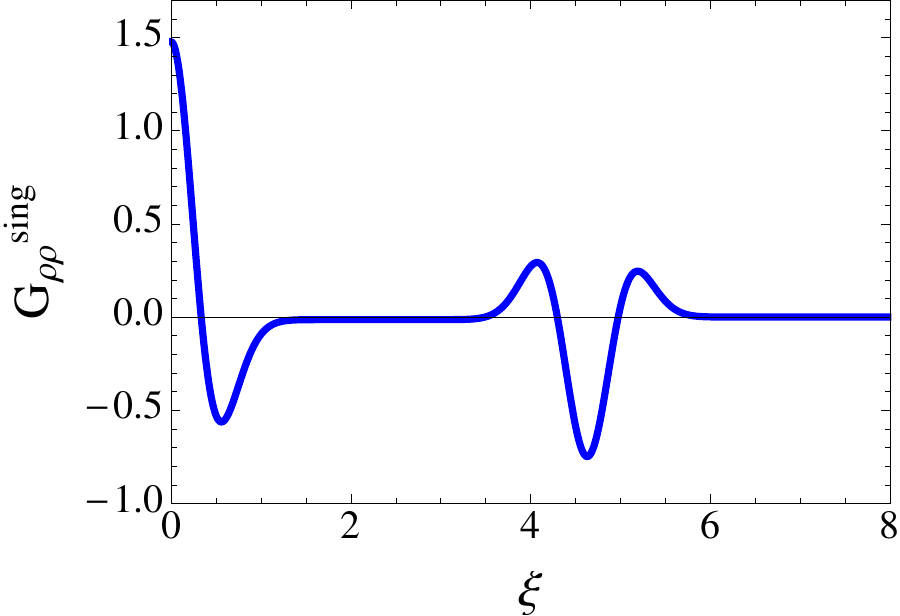}
  \caption{The singular part of the correlator $G_{\rho\rho}(\xi;{\tau_f},\tau)$ with $v_s^2=1/3$ and $\ln({\tau_f}/\tau)=4$. 
    The function is smeared by a Gaussian of variance $\sigma^2=0.1$ in order to
    show the nature of the singularities.}
  \label{fig:G_sing}
\end{figure}


Fluctuations in the local temperature and flow velocity fields give rise to a nontrivial 2-particle correlation when the fluid elements freeze-out to free-streaming hadrons.  In momentum-space rapidity this correlation is
\begin{equation}
\left\langle \frac{d N(\eta_2)}{d\eta} \frac{d N(\eta_1)}{d\eta}  - \left\langle \frac{dN}{d\eta} \right\rangle^2 \right\rangle {\left\langle  \frac{dN}{d\eta}\right\rangle}^{-1} = 
\frac{15 d_s}{\pi^4 N_{\rm eff}}\,\frac{1}{ T_f\tau_f}
\left(\frac{T_0}{T_f}\right)^2  \left(\frac{\eta}{s}\right)_0 K(\Delta\eta)
\label{eq:dNdN/N}
\end{equation}
where $N_{\rm eff}$ is the effective number of bosonic degrees of freedom at the initial time $\tau_0$ and temperature $T_0$ and $d_s$ is the spin/isospin degeneracy of the hadron species.  The correlation is directly proportional to the ratio of shear viscosity to entropy density, $(\eta/s)_0$, assumed here to be temperature independent.  The function $K(\Delta \eta)$ can be computed numerically for a given hadron mass.  It is shown in Fig. 3 for the choice $T_0=600$ MeV, $T_f=150$ MeV, $\tau_f=10$ fm, and $N_{\rm eff}=47.5$.  Folding the fluctuations with the thermal distribution function smooths out the singularities.  Incorporation of viscosity in the expansion dynamics and finite range correlations fill-in the dip at $\Delta \eta=1.5$ but otherwise does not much affect the shape or magnitude of $K$.   It reminds one very much of the near-side ridge \cite{STAR,PHOBOS,CMS}.  
\begin{figure}[b]
  \centering  
\includegraphics[width=0.85\linewidth]{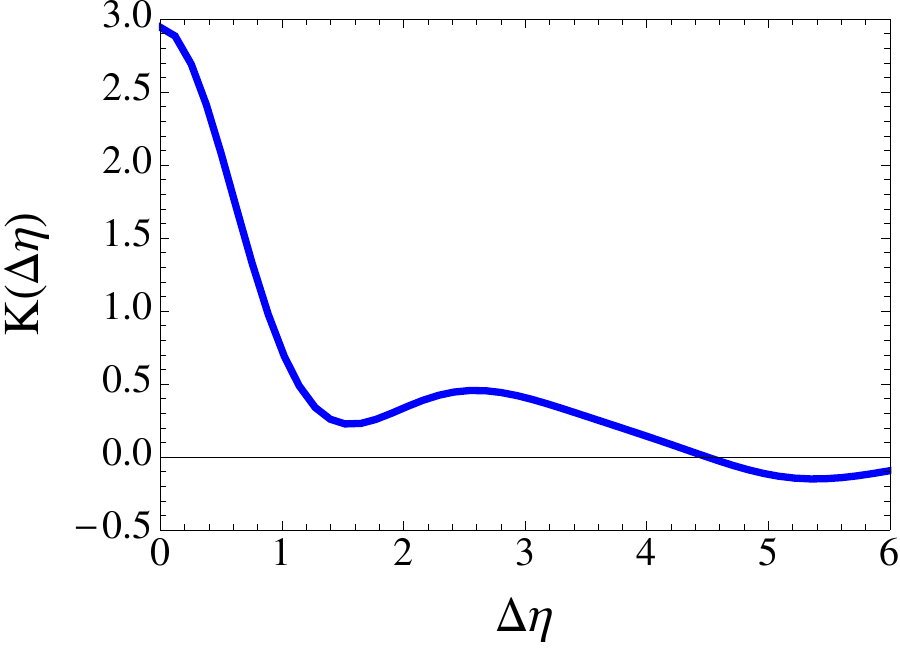}
  \caption[]{The correlation function $K(\Delta\eta)$ in the $dN/d\eta$ fluctuations.}
\label{fig:visc}
\end{figure}


\section{Conclusion}

In this work we have derived the general theory of relativistic hydrodynamic fluctuations and showed how to apply them to high energy heavy ion collisions.  We illustrated the general procedure with the simple example of boost-invariant hydrodynamics.  This example suggests long-range correlations in rapidity caused by hydrodynamic fluctuations in temperature and flow velocity.  Allowing for transverse expansion should also give rise to correlations in azimuth \cite{Staig}.  The magnitude of the correlations are directly proportional to the viscosities, and the range and shape are sensitive to the speed of sound of the medium.  Many more details and references are to be found in an upcoming publication.

To fully explore the implications of the ideas and formulas presented here and for detailed comparison to data requires a full 3+1 dimensional relativistic dissipative hydrodynamic code such as in \cite{Gale}.  It will also be interesting to explore the implications for a critical point in the QCD phase diagrams where fluctuations should be enhanced \cite{criticalpt}.  

Our conclusion is that fluctuations are interesting and can provide important information on transport coefficients.  We are learning, we are enjoying, and there is plenty of work ahead for both theorists and experimentalists.

Finally I congratulate Johann Rafelski on his 60th birthday!  I have known Jan for more than 30 years, but we have collaborated on only one project. With Berndt M\"uller we put together an annotated reprint collection entitled {\it Quark-Gluon Plasma: Theoretical Foundations}.  We assembled what we considered to be the pioneering papers in this field in 13 different categories.  The oldest paper reprinted was Fermi's 1950 article.  The newest papers were from 1992-93 after which essentially all articles became easily accessible and searchable on the preprint archive.  The collection is over 800 pages, and was published by Elsevier in 2003. I highly recommend it for newcomers for a primer on the history of our field.  Thank you, Jan, for having the original idea, and for including me in that project!  

\section*{Acknowledgements}

The work of J. K. was supported in part by the U.S. DOE Grant No. DE-FG02-87ER40328, the work of B. M. in part by the U.S. DOE Grant No. DE-FG02-05ER41367, and the work of M. S. in part by the U.S. DOE Grant No. DE-FG02-01ER41195.


\clearpage

\begin{thebibliography}{99}

\bibitem{WMAP}
D. Larson, {\it et al.} [WMAP], Astrophys. J. Suppl. {\bf 192}, 16 (2011);  E. Komatsu, {\it et al.} [WMAP], Astrophys. J. Suppl. {\bf 192}, 18 (2011).

\bibitem{jets}
For an overview of the first results of Pb-Pb collisions at the LHC see: {\it Proceedings of the 22nd International Conference on Ultra-Relativistic Nucleus–Nucleus Collisions}, J. Phys. G {\bf 38}, No. 12 (2011), eds. Y. Schutz and U. A. Wiedemann.
 
\bibitem{Dodelson} 
  S. Dodelson,  {\it Modern Cosmology}, Academic Press, San Diego (2003).

\bibitem{Donev}
A. Donev, E. Vanden-Eijnden, A. L. Garcia, and J. B. Bell, Comm. App. Math. Comp. Sci. {\bf 5}, 149 (2010).

\bibitem{Eggers}
J. Eggers, Phys. Rev. Lett. {\bf 89}, 084502 (2002).

\bibitem{Kang}
W. Kang and U. Landman, Phys. Rev. Lett. {\bf 98}, 064504 (2007).

\bibitem{fluid1} 
  S. Weinberg, 
  {\it Gravitation and Cosmology}, Wiley, New York (1972).

\bibitem{fluid2} 
  L. D. Landau and E. M. Lifshitz, 
  {\it Fluid Mechanics}, Pergamon Press, Oxford (1987).

\bibitem{statphys2} 
L. D. Landau and E. M. Lifshitz, 
{\it Statistical Physics: Part 2}, Pergamon Press, Oxford (1980).

\bibitem{STAR}
STAR Collaboration, Phys. Rev. C {\bf 80}, 064912 (2009).

\bibitem{PHOBOS}
PHOBOS Collaboration, Phys. Rev. Lett. {\bf 104}, 062301 (2010).

\bibitem{CMS}
CMS Collaboration, JHEP {\bf 7}, 76 (2011).

\bibitem{Staig}
 P. Staig and E. Shuryak,  Phys. Rev. C {\bf 84}, 034908 (2011); {\it ibid}. 044912 (2011).

\bibitem{Gale}
B. Schenke, S. Jeon, and C. Gale, Phys. Rev. C {\bf 82}, 014903 (2010).

\bibitem{criticalpt}
J. I. Kapusta, Phys. Rev. C {\bf 81}, 055201 (2010).

\end{thebibliography}
\end{document}